\documentclass[showpacs,preprintnumbers,
amsmath,amssymb,aps,prd]{revtex4}
\usepackage{amsfonts}
\usepackage{amsfonts,graphics,epsfig,subfigure}

\begin{document}

\title{Universal ratios of critical physical quantities of charged AdS black holes}
\author{Jie-Xiong Mo, Gu-Qiang Li, Xiao-Bao Xu}
 \affiliation{Institute of Theoretical Physics, Lingnan Normal University, Zhanjiang, 524048, Guangdong, China}

\begin{abstract}
 We investigate the ratios of critical physical quantities related to the $T-S$ criticality of charged AdS black holes. It is shown that the ratio $\frac{T_cS_c}{Q_c}$ is universal while $\frac{T_cr_c}{Q_c}$ is not. This finding is quite interesting considering the former observation that both the $T-S$ graph and $T-r_+$ graph exhibit reverse van der Waals behavior. It is also worth noting that the value of $\frac{T_cS_c}{Q_c}$ differs from that of $\frac{P_cv_c}{T_c}$ for $P-V$ criticality. Moreover, we discuss universal ratios for the $P-V$ criticality and $Q-\Phi$ criticality. We successfully interpret the former finding that the ratio $\frac{\Phi_cQ_c}{T_c}$ is not universal and construct two universal ratios for the $Q-\Phi$ criticality instead. To the best of our knowledge, we are the first to introduce the dimensional analysis technique to study the ratios of critical physical quantities. It is expected that this technique can be generalized to probe the universal ratios for $Y-X$ criticality in future research.
\end{abstract}
\keywords{universal ratio\;charged AdS black holes\;$T-S$ criticality}
 \pacs{04.70.Dy, 04.70.-s} \maketitle

\section{Introduction}
     Phase transitions and critical phenomena of charged AdS black holes have long been fascinating topics ever since the pioneer work of Chamblin et al.~\cite{Chamblin1,Chamblin2} disclosed the amazing close relation between charged AdS black holes and van der Waals liquid-gas system. Regarding the cosmological constant $\Lambda$ as a variable \cite{Caldarelli}-\cite{Cvetic} and identifying it as the thermodynamic pressure, the $P-V$ criticality revolution initiated by Kubiz\v{n}\'{a}k and Mann~\cite{Kubiznak} further enhances this relation. One can read the most recent review \cite{Kubiznak2} and references therein to gain an elegant and unified picture.

     Recently, $T-S$ criticality of AdS black holes also attracts much attention. Ref. \cite{Spallucci} first presented the exact construction of the Maxwell equal area law in the $T-S$ plane of charged AdS black hole. It was argued that the behavior of the $T-S$ plane is dual to that of the $P-V$ plane and this duality is analogous to the $T$-duality of string theory. This intriguing work was soon generalized to entanglement entropy-temperature plane creatively \cite{Johnson, Nguyen, Caceres}. From then on, the phase structure of various black holes have been investigated in either the $T-S$ plane or entanglement entropy-temperature plane, including five-dimensional RN-AdS black holes~\cite{Nguyen, zengxiaoxiong4}, Born-Infeld AdS black holes~\cite{zengxiaoxiong3}, quintessence black holes~\cite{zengxiaoxiong2}, $f(R)$ AdS black holes~\cite{xiong7}, Gauss-Bonnet black holes~\cite{zhaoliu, zengxiaoxiong5}, dilaton black holes~\cite{xiong8}, dimensionally continued AdS black holes~\cite{kuangxiaomei} and black holes in massive gravity~\cite{zengxiaoxiong1}.

     However, the discussion of universal ratios of critical quantities related to the $T-S$ criticality was unfortunately missed in all these literatures (they only calculated critical quantities). Probing the universal ratios of critical quantities is of great physical significance. In classical thermodynamics, the ratio $\frac{P_cv_c}{kT_c}$ is a universal number for all van der Waals fluids. It was shown in the pioneering work of $P-V$ criticality~\cite{Kubiznak} that four-dimensional RN-AdS black holes with arbitrary charge share exact the same universal ratio as the van der Waals fluids. Namely, $\frac{P_cv_c}{T_c}=\frac{3}{8}$. Ref.~\cite{Gunasekaran} further showed that this ratio was worked out to be $\frac{2d-5}{4d-8}$ for the $d$-dimensional charged black holes and it is also independent of the parameter $q$ which is related to the charge of black holes.

     Considering the duality between the behavior of the $T-S$ plane and that of the $P-V$ plane, we believe there also exist universal ratios of critical quantities related to the $T-S$ criticality. In this paper, we will attack this problem under the background of $d$-dimensional charged AdS black hole spacetime. The $T-S$ criticality of $d$-dimensional charged black holes has not been covered in literature yet to the best of our knowledge (As stated in the second paragraph, only four and five-dimensional charged black holes were covered. Ref. \cite{Belhaj} studied the coexistence line, the Maxwell area law and the heat engine of $d$-dimensional black holes but it did not study their critical quantities for the $T-S$ criticality). On the other hand, Ref. \cite{mayubo} recently investigated the $Q-\Phi$ criticality of $d$-dimensional charged black holes and argued that the ratio $\frac{\Phi_cQ_c}{T_c}$ is not universal. To explain this phenomenon, we will also carry out some discussion on the universal ratios for $P-V$ criticality and $Q-\Phi$ criticality in this paper. And we will pave the way to construct universal ratios for the $Q-\Phi$ criticality. We believe that the technique we used in this paper can be generalized to probe the universal ratios for $Y-X$ criticality in future research.

    The organization of this paper is as follows. In Sec.\ref{Sec2} we will investigate the $T-S$ criticality of $d$-dimensional charged AdS black holes and search for universal ratios of critical quantities related to the $T-S$ criticality. Sec.\ref{Sec3} will be devoted to the discussions on the universal ratios for $P-V$ criticality and $Q-\Phi$ criticality. In the end, a brief conclusion will be presented in Sec.\ref {Sec4}.

\section{Universal ratios for $T-S$ criticality of charged AdS black holes}
\label{Sec2}
\subsection{Setup}
The bulk action of the $d$-dimensional ($d>3$) charged AdS black hole reads
\begin{equation}
I_{EM}=-\frac{1}{16\pi}\int_M d^dx\sqrt{-g}\left(R-F^2-2\Lambda\right),\label{1}
\end{equation}%
where $\Lambda$ is the cosmological constant which is related to the characteristic length scale $l$ through $\Lambda=-\frac{(d-1)(d-2)}{2l^2}$.

The corresponding solution of this action have been reviewed in Ref.~\cite{Gunasekaran} as
\begin{eqnarray}
ds^2&=&-fdt^2+\frac{dr^2}{f}+r^2d\Omega_{d-2}^2, \nonumber
\\
F&=&dA, \; \; A=-\sqrt{\frac{d-2}{2(d-3)}}\frac{q}{r^{d-3}}dt,\label{2}
\end{eqnarray}%
where
\begin{equation}
f=1-\frac{m}{r^{d-3}}+\frac{q^2}{r^{2(d-3)}}+\frac{r^2}{l^2}.\label{3}
\end{equation}%
The ADM mass and the electric charge of the black hole has been identified as ~\cite{Chamblin1}
\begin{eqnarray}
M&=&\frac{\omega_{d-2}(d-2)}{16\pi}m,\label{4}
\\
Q&=&\frac{\omega_{d-2}\sqrt{2(d-2)(d-3)}}{8\pi}q,\label{5}
\end{eqnarray}%
where $\omega_d=\frac{2\pi^{\frac{d+1}{2}}}{\Gamma(\frac{d+1}{2})}$ denotes the volume of the unit $d$-sphere.

The corresponding Hawking temperature, entropy and electric potential have been reviewed in Ref.~\cite{Gunasekaran} as
\begin{eqnarray}
T&=&\frac{f'(r_+)}{4\pi}=\frac{d-3}{4\pi r_+}\left(1-\frac{q^2}{r_+^{2(d-3)}}+\frac{d-1}{d-3} \frac{r_+^2}{l^2}\right),\label{6}
\\
S&=&\frac{\omega_{d-2}r_+^{d-2}}{4}, \label{7}
\\
\Phi&=&\sqrt{\frac{d-2}{2(d-3)}}\frac{q}{r_+^{d-3}}. \label{8}
\end{eqnarray}%
\subsection{Universal ratios for $T-S$ criticality}

The possible critical point of $T-S$ graph is determined by the following equations
\begin{eqnarray}
\left(\frac{\partial T}{\partial S}\right)_{q=q_c, S=S_c}&=&0,\label{9}
\\
\left(\frac{\partial^2 T}{\partial S^2}\right)_{q=q_c, S=S_c}&=&0,\label{10}
\end{eqnarray}%
where the subscript "c" denotes the values of physical quantities at the critical point.

The analytic expressions for the critical quantities can be obtained via tedious calculation as
\begin{eqnarray}
S_c&=&\frac{\omega_{d-2}}{4}\left[\frac{(d-3)l}{\sqrt{(d-2)(d-1)}}\right]^{d-2},\label{11} \\
q_c&=&[(d-3)l]^{(d-3)}\sqrt{\frac{1}{(2d-5)(d-2)^{d-2}(d-1)^{d-3}}},\label{12} \\
T_c&=&\frac{(d-3)\sqrt{(d-2)(d-1)}}{(2d-5)l\pi}.\label{13}
\end{eqnarray}%
One can further obtain the critical charge $Q_c$ and the critical horizon radius $r_c$as
\begin{eqnarray}
Q_c&=&\frac{(d-3)^{\frac{2d-5}{2}}l^{d-3}\omega_{d-2}}{4\sqrt{4d-10}[(d-2)(d-1)]^{\frac{d-3}{2}}\pi},\label{14} \\
r_c&=&\frac{(d-3)l}{\sqrt{(d-1)(d-2)}}.\label{15}
\end{eqnarray}%
From Eqs. (\ref{11})-(\ref{15}), one can see clearly that these critical quantities depend on both the dimensionality $d$ and the characteristic length scale $l$.

With these critical quantities on hand, we can calculate various ratios of these critical physical quantities and check whether they are universal. The results are listed as
\begin{eqnarray}
\frac{T_cS_c}{q_c}&=&\frac{(d-3)^2\omega_{d-2}}{4\pi}\sqrt{\frac{d-2}{2d-5}},\label{16} \\
\frac{T_cS_c}{Q_c}&=&\frac{\sqrt{2}(d-3)^{3/2}}{\sqrt{2d-5}},\label{17} \\
\frac{T_cr_c}{q_c}&=&\frac{(d-3)^{5-d}[(d-1)(d-2)]^{\frac{d-2}{2}}}{\pi l^{d-3}\sqrt{(d-1)(2d-5)}},\label{18} \\
\frac{T_cr_c}{Q_c}&=&\frac{4\sqrt{2}(d-3)^{\frac{9-2d}{2}}[(d-1)(d-2)]^{\frac{d-3}{2}}}{\omega_{d-2}l^{d-3}\sqrt{2d-5}}.\label{19} \\
\end{eqnarray}%
It can be seen clearly that the ratios $\frac{T_cS_c}{q_c}$ and $\frac{T_cS_c}{Q_c}$ only depend on the dimensionality $d$ while the ratios $\frac{T_cr_c}{q_c}$ and $\frac{T_cr_c}{Q_c}$ depend on both the dimensionality $d$ and the characteristic length scale $l$. In this sense, the ratios $\frac{T_cS_c}{q_c}$ and $\frac{T_cS_c}{Q_c}$ are universal (independent of $l$) while the other two ratios are not. This finding is quite interesting considering the former observation \cite{xiong7} that both the $T-S$ graph and $T-r_+$ graph exhibit reverse van der Waals behavior. It is also worth noting that the universal ratio $\frac{T_cS_c}{Q_c}$ for $T-S$ criticality differs from that for $P-V$ criticality~\cite{Gunasekaran}. The latter one reads $\frac{P_cv_c}{T_c}=\frac{2d-5}{4d-8}$~\cite{Gunasekaran}.

\section{Discussions on universal ratios for $P-V$ criticality and $Q-\Phi$ criticality of charged AdS black holes}
\label{Sec3}
\subsection{Universal ratios for $P-V$ criticality}
$P-V$ criticality of $d$-dimensional ($d>3$) charged AdS black hole has been investigated elegantly in Ref.~\cite{Gunasekaran}. The critical quantities have been obtained as~\cite{Gunasekaran}
\begin{eqnarray}
v_c&=&\frac{1}{\kappa}\left[q^2(d-2)(2d-5)\right]^{1/[2(d-3)]},\label{20} \\
T_c&=&\frac{(d-3)^2}{\pi \kappa v_c(2d-5)},\label{21} \\
P_c&=&\frac{(d-3)^2}{16\pi \kappa^2 v_c^2}.\label{22}
\end{eqnarray}%
The specific volume $v$ is related to the horizon radius $r_+$ through $r_+=\kappa v$, where $\kappa=\frac{d-2}{4}$. It was shown that the ratio $\frac{P_cv_c}{T_c}$ is universal (independent of the parameter $q$)~\cite{Gunasekaran}.

Here, we would like to discuss on another two ratios which are also related to the $P-V$ criticality of charged AdS black holes. The results are listed as
\begin{eqnarray}
\frac{P_cr_c}{T_c}&=&\frac{2d-5}{16},\label{23} \\
\frac{P_cV_c}{T_c}&=&\frac{(2d-5)[(d-2)(2d-5)q^2]^{\frac{d-2}{2d-6}}\omega_{d-2}}{16(d-1)}.\label{24}
\end{eqnarray}%
Note that $r_c$ and $V_c$ is obtained via Eq. (\ref{20}) utilizing the relations among these three quantities, which read $r_+=\kappa v$ and $V=\frac{\omega_{d-2}r_+^{d-1}}{d-1}$.

It can be seen that the ratio $\frac{P_cr_c}{T_c}$ is also universal since it does not depend on the parameter $q$. However, the ratio $\frac{P_cV_c}{T_c}$ is not universal and depends on both the dimensionality $d$ and the parameter $q$. This finding seems to support the former observation that the specific volume $v$ rather than the thermodynamic volume $V$ plays the crucial role in the $P-V$ criticality of charged AdS black holes \cite{Kubiznak}.

\subsection{Universal ratios for $Q-\Phi$ criticality}
Ref.~\cite{mayubo} studied $Q-\Phi$ criticality of $(n+1)$-dimensional (Note that $n=d-1$.) charged AdS black hole. The critical quantities have been presented as~\cite{mayubo}
\begin{eqnarray}
\Phi_c&=&\frac{\pi^{n/2-1}}{4\Gamma(n/2)}\sqrt{\frac{n-1}{2n-3}},\label{25} \\
T_c&=&\frac{n-2}{\pi (2n-3)}(-2\Lambda)^{1/2},\label{26} \\
Q_c&=&[-\frac{(n-2)^2}{2\Lambda}]^{(n-2)/2}[(n-1)(2n-3)]^{-1/2}.\label{27}
\end{eqnarray}%
And the ratio $\frac{\Phi_cQ_c}{T_c}$ was obtained as
\begin{equation}
\rho_c=\frac{\sqrt{-\Lambda}2^{-\frac{n}{2}-\frac{3}{2}}(n-1)\pi^{n/2}[-\frac{(n-2)^2}{\Lambda}]^{n/2}}{(n-2)^3\sqrt{\frac{n-1}{2n-3}}\sqrt{2n^2-5n+3}\Gamma\left(\frac{n}{2}\right)}.\label{28}
\end{equation}%
Based on the above equation, the authors of Ref.~\cite{mayubo} argued that this critical ratio depends on both $\Lambda$ and $n$, which is different from the $P-V$ criticality.

However, $(V,P)$, $(\Phi,Q)$ and $(T,S)$ are conjugate pairs determined by the first law of black hole thermodynamics $dM=TdS+\Phi dQ+VdP$. In this sense, they should behave similarly. In fact, it has been shown that the duality of descriptions in $P-V$ plane and $T-S$ plane is analogous to the $T$-duality of string theory \cite{Spallucci}. So we believe that there also exist universal ratios for $Q-\Phi$ criticality, just as for $T-S$ criticality and $P-V$ criticality. The phenomenon that the ratio $\frac{\Phi_cQ_c}{T_c}$ fails to show the universal property may be attributed to the fact that this ratio is not dimensionless. One can easily show that this ratio has dimension $[$length$]^{d-2}$ via dimensional analysis. The same technique can also be used to explain why the ratio $\frac{P_cV_c}{T_c}$ is not universal.

 Here, we construct two ratios for the $Q-\Phi$ criticality of charged AdS black holes as follows
\begin{eqnarray}
\frac{\Phi_cQ_c^{\frac{1}{2-n}}}{T_c}&=&\frac{[(2n-3)(n-1)]^{\frac{n-1}{2(n-2)}}\pi^{n/2}}{4(n-2)^2\Gamma\left(\frac{n}{2}\right)},\label{29} \\
\frac{\Phi_cQ_c}{T_c^{2-n}}&=&\frac{(n-2)^{2n-4}(2n-3)^{1-n}\pi^{1-n/2}}{4\Gamma\left(\frac{n}{2}\right)}.\label{30} \\
\end{eqnarray}%
These two ratios only depend on $n$, which is related to the dimensionality $d$ via $d=n+1$. In this sense, they are universal.

\section{Conclusions}
\label{Sec4}
    In this paper, we investigate the ratios of critical physical quantities related to the $T-S$ criticality of charged AdS black holes. It is shown that the ratios $\frac{T_cS_c}{q_c}$ and $\frac{T_cS_c}{Q_c}$ only depend on the dimensionality $d$ while the ratios $\frac{T_cr_c}{q_c}$ and $\frac{T_cr_c}{Q_c}$ depend on both the dimensionality $d$ and the characteristic length scale $l$. In this sense, the ratios $\frac{T_cS_c}{q_c}$ and $\frac{T_cS_c}{Q_c}$ are universal (independent of $l$) while the other two ratios are not. This finding is quite interesting considering the former observation that both the $T-S$ graph and $T-r_+$ graph exhibit reverse van der Waals behavior. It is also worth noting that the universal ratio $\frac{T_cS_c}{Q_c}$ for $T-S$ criticality differs from $\frac{P_cv_c}{T_c}$ for $P-V$ criticality~\cite{Gunasekaran}. The value of the latter one was shown to be $\frac{2d-5}{4d-8}$~\cite{Gunasekaran}.

    Moreover, we discuss on universal ratios for $P-V$ criticality and $Q-\Phi$ criticality of charged AdS black holes. It is shown that the ratio $\frac{P_cr_c}{T_c}$ is also universal while the ratio $\frac{P_cV_c}{T_c}$ is not. This finding further supports the former observation that the specific volume $v$ rather than the thermodynamic volume $V$ plays the crucial role in the $P-V$ criticality of charged AdS black holes \cite{Kubiznak}. Recently, Ref. \cite{mayubo} showed that the ratio $\frac{\Phi_cQ_c}{T_c}$ fails to show the universal property. We attribute this phenomenon to the fact that the ratio $\frac{\Phi_cQ_c}{T_c}$ is not dimensionless. And we successfully construct two universal ratios for the $Q-\Phi$ criticality of charged AdS black holes. To the best of our knowledge, we are the first to introduce the dimensional analysis technique to study the ratios of critical physical quantities. It is expected that this technique can be generalized to probe the universal ratios for $Y-X$ criticality in future research. Furthermore, it is expected that there also exists universal ratios for the Hawking temperature-Entanglement entropy criticality. And we will return to this issue in the near future.
    
    Last but the least, our paper disclosed a variety of interesting phenomena whose physical interpretation deserves further investigation. And we would like to call for the attention to the following problems. Firstly, both the $T-S$ graph and $T-r_+$ graph exhibit reverse van der Waals behavior. Why the ratio $\frac{T_cS_c}{Q_c}$ is universal while $\frac{T_cr_c}{Q_c}$ is not? Secondly, why the ratio $\frac{P_cr_c}{T_c}$ rather than the ratio $\frac{P_cV_c}{T_c}$ is universal? This is not straightforward considering the fact that it is $V$ rather than $r_+$ enters the first law of black hole thermodynamics. Thirdly, the physical interpretation why the ratio $\frac{\Phi_cQ_c}{T_c}$ is not universal is also intriguing. The above three problems we proposed in this paper seem separate at the first sight, but the answer to these questions may help build a unified picture of black hole thermodynamics. It is hoped that the deep physics behind the above phenomena can be disclosed one day.

 \section*{Acknowledgements}
The authors are supported by National Natural Science Foundation of China (Grant No.11605082), and in part supported by Natural Science Foundation of Guangdong Province, China (Grant Nos.2016A030310363, 2016A030307051, 2015A030313789)

\end{document}